\documentclass{appolb}
\usepackage{epsfig}

\begin{document}
\date{\today}
\pagestyle{plain}
\title{Controlling diffusive transport in confined geometries}
\author{P.S. Burada \address{Max-Planck Institute f\"ur Physik komplexer Systeme,
  N\"othnitzer Str. 38, 01187 Dresden, Germany}
\and G. Schmid, Y. Li, P. H\"anggi \address{Institut f\"ur Physik,
  Universit\"at Augsburg,
  Universit\"atsstr. 1,
  86135 Augsburg, Germany}
}

\maketitle
\begin{abstract}
We analyze the diffusive transport of Brownian particles in narrow 
channels with periodically varying cross-section. 
The geometrical confinements lead to entropic barriers, the particle has to overcome in 
order to proceed in transport direction. 
The transport characteristics exhibit peculiar behaviors which
are in contrast to what is observed for the transport in 
potentials with purely energetic barriers.
By adjusting the geometric parameters of the channel one can effectively
tune the transport and diffusion properties. A prominent example is the 
maximized enhancement of diffusion for particular channel parameters.
The understanding of the role of channel-shape
provides the possibility for a design of stylized 
channels wherein the quality of the transport can be efficiently optimized.
\end{abstract}

\section {Introduction}

Diffusion process in narrow confined systems
exhibits peculiar properties which are radically different from what we
generally observe in free systems, i.e., in the absence of any
geometrical restrictions \cite{Burada_CPC}.
If we consider a Brownian particle which is moving in a potential
energy landscape or in a confined geometry where the geometrical 
confinement may regulate or control the diffusion process, 
the diffusion varies significantly from the free 
case \cite{Reimann_PRL,Lindner_FNL,Reguera_PRL}.
Depending on the conditions imposed, the diffusion coefficient can be 
larger or smaller than the bulk diffusion coefficient.
Brownian particles when moving in a confined geometry
undergo a constrained diffusive motion.
This feature of constrained motion is ubiquitous in
ion channels, nanopores, zeolites, and for processes occurring
at sub-cellular level \cite{Reguera_PRL,hille,liu, siwy,berzhkovski,zeolites,Chou,
kettner,muller,Austin,Nixon,Chang,Kosinska}.
The uneven shape of these structures regulates the transport
of particles yielding important effects exhibiting peculiar
properties \cite{Reguera_PRL,Austin,Nixon,Chang}.
The results have implications in processes such as
catalysis, osmosis and particle separation
\cite{hille,liu,siwy,berzhkovski,kettner,muller}
and on the noise-induced transport in 
periodic potential landscapes that lack reflection symmetry, 
such as ratchet systems \cite{revmodphys,BM,PT,RH}. 

Nowadays, artificial, i.e., synthetic, nanopores are
available \cite{siwy,Keyser-RSIns,Han} to characterize 
the transport characteristics 
of ionic species like $\mathrm{K}^{+}, \mathrm{Na}^{+}$,   
$\mathrm{Cl}^{-}$ \cite{siwy,Kosinska} and of
macro molecules, 
like for example, DNA or RNA \cite{Han,Keyser,Aksimentiev,Heng_1,Heng_BPJ}.
With latest technology, these synthetic nanopores can be made
upon choice, allowing for effective control of the diameter 
and the shape of the nanoporous structure \cite{siwy,Keyser-RSIns,Han}.
In recent years, it has been of great interest to reveal the sequence 
and structural analysis of DNA and RNA molecules by passing them through 
nanopores \cite{Keyser,Aksimentiev,Heng_1,Heng_BPJ,Gerland,Bundschuh}.
When a double stranded DNA molecule passes through the charged 
nanopore, each base pair exhibits its own distinct electronic signature
because each base pair is structurally and chemically different.
While present conventional methods to find out this sequence would 
take several months and are expensive, 
one could find the sequence of a human genome
in a matter of hours at a potentially low cost 
using this new technology. In addition, charging of the inner tube walls could lead to 
rectification of the motion of ions \cite{siwy,Kosinska,Kosinska_APP}.  

It has been found that the separation of macromolecule fragments 
and ions in narrow channels \cite{Austin,Nixon,Chang,Han,Aksimentiev} 
is largely influenced by their shape alone. The characteristic behavior of ionic
currents of these molecules or individual species can effectively
change if the pore diameter changes\cite{Aksimentiev,Heng_BPJ,Brun,Kasianowicz,Choi}. 
The subject of this paper is to systematically illustrate the role of the channel properties
on the transport characteristics for Brownian particles passing through the channels.  

The paper is organized in the following way.
In section II we introduce the model system, the dynamics and
discuss the complications involved in solving the full problem.
The reduction of dimensionality, or in other words the simplification,
and the analytical treatment of the problem is discussed in Section III.
In Section IV we introduce the simulation techniques to solve the full problem
numerically. The main findings on the transport characteristics are discussed in Section V.
In Section VI we present the main conclusions.

\section{Modeling}

\begin{figure}[t]
 \centering
 \includegraphics{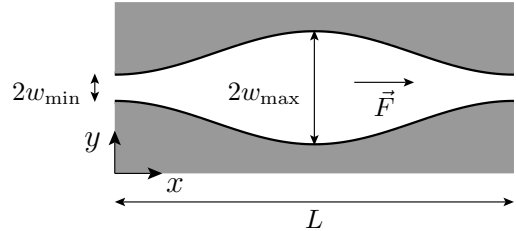}
 \caption{Schematic representation of a single channel
   with the periodicity $L$, bottleneck half-width $w_{\mathrm{min}}$
   and maximal half-width $w_{\mathrm{max}}$.
   The shape of the structure is defined by Eq.~(\ref{eq:geometry}).
   The constant applied bias $F$ acting only along the length of the channel.}
 \label{fig:geometry}
\end{figure}

We investigate the transport of Brownian particles through straight 
channels with periodically varying cross-section. 
In the following, we mainly focus on $2D$ - channels 
(see Fig.~\ref{fig:geometry}), although a very similar line of reasoning could be applied for  
pore structures in $3D$. 
For a model system we consider a channel structure 
whose shape is defined by the periodic boundary function $\omega(x)$ 
with the periodicity $L$, i.e. $\omega(x+L) = \omega(x)$. 
The diffusive motion of the Brownian particle is then confined by 
the upper boundary $\omega(x)$ and the lower boundary 
$-\omega(x)$. In particular, we choose:
\begin{equation}
\label{eq:boundaryfct}
\omega(x) = A \sin( 2\pi \frac{x}{L} ) + B\, ,
\end{equation}
with the two channel parameters $A$ and $B$ and where the function $\omega(x)$ 
is thought to be the first terms of the 
Fourier series of a more complex boundary function. Due to the symmetry 
with respect to the $x$-axis the boundary function 
could be given in terms of the maximum half-width of the channel 
$w_\mathrm{max} = ( A + B )$ and the aspect ratio of maximum and 
minimum width $\epsilon = ( B - A ) / ( B + A )$, i.e. 
\begin{equation}
\label{eq:geometry} \omega(x) =
\frac{w_{\mathrm{max}}}{2}\,(1-\epsilon)\, \left\{\sin(2\pi \frac{x}{L}) +
\frac{1+\epsilon}{1-\epsilon} \right\} \, .
\end{equation}

Transport through the considered confined geometry may be caused by
different particle concentrations maintained at the ends of the
channel, or by the application of external forces acting on the
particles. Here we will exclusively consider the case of force
driven transport. The external driving force is denoted by
$\vec{F} = F \vec{e}_x$ which points into the direction of the
channel axis. Then the dynamics of a suspended Brownian particle
is described by means of Langevin equation, in the over damped
limit \cite{Purcell}, which reads
\begin{equation}
  \label{eq:langevin}
  \gamma\, \frac{\mathrm{d}\vec{x}}{\mathrm{d} t} = \vec{F} +
  \sqrt{\gamma \, k_\mathrm{B}\,T}\, \vec{\xi}(t)\, ,
\end{equation}
where $t$ is time, $\vec x$ the position vector of the particle,
$\gamma$ its friction coefficient, $k_{\mathrm{B}}$ the Boltzmann
constant and $T$ the temperature. The thermal fluctuating forces
which the surrounding fluid exerts on the particle are modeled by
zero-mean Gaussian white noise $\vec{\xi}(t)$, obeying the
fluctuation-dissipation relation $\langle \xi_{i}(t)\,\xi_{j}(t')
\rangle = 2\, \delta_{ij}\, \delta(t-t')$ for $i,j = x,y$. In
addition to Eq.~(\ref{eq:langevin}), the full problem is set up by
imposing reflecting boundary conditions at the channel walls.
For the sake of simplicity, we measure 
lengths by the periodicity of the channel $L$, 
time by $\tau=\gamma {L}^2/(k_{\mathrm{B}}T_{\mathrm{R}})$ 
which is the corresponding characteristic diffusion time 
at an arbitrary but irrelevant reference temperature $T_{\mathrm{R}}$,
and force by $F_{\mathrm{R}}=k_{\mathrm{B}}T_{\mathrm{R}}/L$.
Consequently, the Langevin equation reads in dimensionless form:
\begin{equation}
  \label{eq:lan-dl}
  \frac{\mathrm{d}\vec{r}}{\mathrm{d} t} =  F \,\vec{e_x}
  +\sqrt{D}\, \vec{\xi}(t)\,,
\end{equation}
where $D = T/T_{\mathrm{R}}$ is the dimensionless temperature, and the boundary function is given by:
\begin{equation}
\label{eq:geometryus} \omega(x) =
\frac{w_{\mathrm{max}}}{2}\,(1-\epsilon)\, \left\{\sin(2\pi x) +
\frac{1+\epsilon}{1-\epsilon} \right\} \, .
\end{equation}

The corresponding Fokker-Planck equation for the time evolution of
the probability distribution $P(\vec r, t)$ takes in dimensionless units the form
\cite{hanggithomas, Risken}:
\begin{equation}
\label{eq:fp}
\frac{\partial P(\vec r, t)}{\partial t} =
- \vec{\nabla} \cdot \left( \vec{F} -
   D\,\vec{\nabla}\right) P(\vec r, t) \, .
\end{equation}
Since, in the present situation, we deal with irregular and impenetratable 
channel walls, the reflection of particles at the boundaries leads to a
vanishing probability current \cite{Burada_CPC}. Solving the full problem in a higher dimensional
space with this irregular boundaries is a difficult task and so far 
there is no analytical method.
Despite the inherent complexity of this problem an
approximate solution can be found  by introducing an effective
one-dimensional kinetic description where geometric constraints and
bottlenecks are considered as entropic barriers
\cite{Reguera_PRL,Jacobs,Zwanzig,Reguera_PRE,Burada_PRE}.

\section{The origin of entropic barriers}
\label{sec:fickjacobs}

The dynamics of the system can be approximatively described by means of
a $1D$ kinetic equation, obtained originally from $3D$ (or $2D$) Smoluchowski equation
by assuming fast equilibration in orthogonal channel direction. This so-called Fick-Jacobs 
equation reads in presence of an applied bias \cite{Reguera_PRL}
\begin{equation}
\label{eq:fickjacobs}
\frac{\partial P(x,t)}{\partial t}=
     \frac{\partial}{\partial x}\left(D(x)\frac{\partial P(x,t)}{\partial x} +
     \frac{D(x)}{D} \frac{\partial A(x)}{\partial x}\, P(x,t) \right) \, ,
\end{equation}
and approximate the full dynamics for $\omega(x) \ll 1$.
In Eq.~(\ref{eq:fickjacobs}), $P(x,t)$ is the probability distribution function along the length 
of the 3D tube or 2D channel, $A(x)$ defines the free energy:
\begin{equation}
\label{eq:freeenergy}
 A(x):= E - TS = -Fx - D\ln s(x)\, , 
\end{equation} 
where $s(x)$ is the dimensionless 
transverse cross section $s(x):= \pi \omega^2(x)$ in three-dimensions, and dimension
less width $s(x):= 2 \omega(x)$ in two-dimensions, where $\omega(x)$ is the radius 
of the pore in $3D$ (or the half-width of the channel in $2D$).
Interestingly, bottlenecks in the channel structure accounts for barriers 
in the potential function $A(x)$ whose height 
scales with the temperature indicating a clear entropic contribution. 
In this terms our model system allows for investigation 
of transport in presence of entropic barriers.

The spatially diffusion coefficient appearing in Eq.~(\ref{eq:fickjacobs}) reads
\begin{equation}
D(x)= \frac{D}{(1+\omega^\prime(x)^2)^\alpha} \, ,
\end{equation}
where $\alpha = 1/3, 1/2$ for two and three dimensions, respectively.
Here, introducing the spatially diffusion coefficient, into the kinetic
equation, improves the accuracy, and extends it's validity to more winding structures 
\cite{Reguera_PRL,Zwanzig,Reguera_PRE,Burada_PRE,Burada_BioSy,Kalinay_PRE,Kalinay_PRE09}.

\section{Transport characteristics}
\label{sec:transchar}

For the transport of Brownian particles the quantities of interest are: 
the mean particle velocity $\langle \dot{x} \rangle$, respectively the 
nonlinear mobility $\mu=\langle \dot{x} \rangle / F$, 
the effective diffusion coefficient $D_\mathrm{eff}$, 
and a measure for the quality of the transport. 
For the latter we take the so-called $Q$-factor, which is the ratio of 
the effective diffusion coefficient and the particle current, reading 
\begin{equation}
Q = \frac{D_\mathrm{eff}}{\langle \dot{x}\rangle}\, .
\label{eq:q}
\end{equation}

\subsection{Analytics} 

After applying the equilibration assumption pointed out in 
Sec.~\ref{sec:fickjacobs}, Eq.~\ref{eq:fickjacobs} could be derived. 
For periodic structures analytical expressions for the above mentioned 
transport characteristics can be determined
by using the mean first passage time approach\cite{Reguera_PRL,Burada_PRE,Burada_BioSy}.
Accordingly, the average particle current is given by
\begin{equation}
  \label{eq:An-current}
  \langle \dot{x} \rangle = \frac{1-e^{-F/D}}
  {\displaystyle \int_{0}^{1} I(z)\,\mathrm{d}z} \, ,
\end{equation}
and the effective diffusion coefficient by
\begin{equation}
  \label{eq:An-diffusion}
  D_{\mathrm{eff}} = D\,\frac{\displaystyle \int_{0}^{1} 
    \displaystyle \int_{x-1}^{x}\,
    \frac{D(z)\,e^{A(x)/D}}{D(x)\,e^{A(z)/D}}\,
    \left[I(z) \right]^2 \, \mathrm{d}x\,\mathrm{d}z}
  {\left[\displaystyle \int_{0}^{1} I(z)\,\mathrm{d}z\,\right]^3} \, ,
\end{equation}
where the integral function reads
\begin{equation}
  \label{eq:integral}
  I(z) = \frac{e^{A(x)/D}}{D(x)} \int_{x-1}^{x}\mathrm{d}y\,e^{-A(y)/D} \, .
\end{equation}       
In the case of an energy barrier, the driving force $F$ and
the temperature $D$ are two independent variables, whereas for
entropic transport, both current and effective diffusion
are controlled by a universal scaling parameter $F/D$ 
\cite{Burada_CPC,Reguera_PRL, Burada_PRE,Burada_BioSy}. 
As, for a given structure, the nonlinear mobility $\mu = \langle \dot{x} \rangle /F$, 
the ratio of effective diffusion $D_\mathrm{eff}$ and bulk diffusion $D$ solely depend on 
the scaling parameter $F/D$, the same applies for the $Q$-factor and Eq.~(\ref{eq:q}) reduces to:
\begin{equation}
Q = \frac{D_\mathrm{eff}/D}{\mu}\, Q_\mathrm{free}\,  ,
\end{equation}
where $Q_\mathrm{free}$ is the $Q$-factor for the biased Brownian 
motion in absence geometrical constraints (free case) which is 
the inverse of the scaling parameter, i.e., $Q_\mathrm{free} =  1 \big/ (F/D)$.

\subsection{Brownian dynamics simulations} 

The approximative transport characteristics described above 
can also be compared with those obtained from precise numerical 
simulations considering the full $2D$ dynamics. 
The mean particle 
current and the effective diffusion coefficient, have 
been corroborated by performing Brownian dynamic simulations
by integrating the Langevin equation Eq.~(\ref{eq:langevin}), 
within the stochastic Euler-algorithm.
Then, the mean velocity in $x$-direction is given by
\begin{equation}
\langle \dot{x}\rangle=\lim_{t\to \infty} \frac{x(t)}{t} \, ,
\label{eq:simcur}
\end{equation}
and the corresponding effective diffusion coefficient reads
\begin{equation}
D_{\mathrm{eff}} = \lim_{t \to \infty} \frac{\langle
x^{2}(t) \rangle - \langle x(t) \rangle^{2}}{2t} \, .
\label{eq:simdiff}
\end{equation}

By comparison of the precise numerical and the analytical results 
for different channel geometries, the 
validity of the equilibration assumption could be 
analyzed in detail \cite{Burada_PRE,Burada_BioSy}.

\section{Geometry controlled transport}

The confinement by the considered channel geometry can be
altered by systematically changing the parameters
$w_{\mathrm{max}}$ and $\epsilon$ in the
geometric function, Eq.~\ref{eq:geometryus}.
Changing these parameters we can
consider two cases: {\itshape constant-ratio-scaling} 
and {\itshape constant-width-scaling}.
In the former case, we modify the maximum width $w_{\mathrm{max}}$ of
the geometry and keep the aspect ratio  $\epsilon$  constant. Consequently, 
by applying the constant-ratio-scaling the minimum width, 
i.e. the width at the bottlenecks, is changing by modifying the maximum width. 
In contrast, in the constant-width-scaling, we keep the maximum width 
$w_\mathrm{max}$ constant and change the aspect ratio $\epsilon$.
In the following we analyze the transport characteristics 
within the two different scaling regimes.

\subsection{Constant-ratio-scaling}
\label{sec:crscaling}

\begin{figure}[t]
 \centering
 \includegraphics{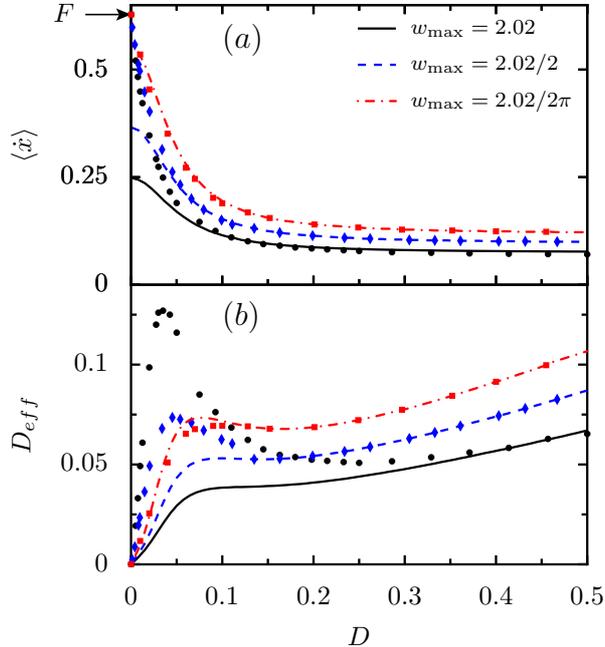}
 \caption{(Color online) Noise (temperature) dependence of the
   average particle current (a) and the effective diffusion 
   coefficient (b) for a symmetric two-dimensional channel 
   with the shape defined by Eq.~(\ref{eq:geometryus}) at the force 
   value $F = 0.628$ and for various values of the maximum 
   half-width ($w_{\mathrm{max}}$) of the geometry and for a 
   constant aspect ratio $\epsilon = 0.01$. The lines correspond
   to the analytic, approximative results given by Eqs.~(\ref{eq:An-current})
   and (\ref{eq:An-diffusion}). The different symbols correspond to
   simulation results, cf. Eqs.~(\ref{eq:simcur}) and
   (\ref{eq:simdiff}). The arrow indicates the particle current for
   the deterministic limit for which the particle current equals $F$.}
 \label{fig:geo1}
 \end{figure}

In constant-ratio-scaling, we fix the ratio of maximum and minimum width. 
Hence, by increasing the maximum width $w_\mathrm{max}$ the channel 
is diluted in orthogonal channel direction. The advantage of this scaling 
is related to the effect, that within the approximative reduced dynamics, 
the height of the entropic barriers is kept fixed while the curvature of 
the effective potential at the minima and maxima varies. 
In Fig.~\ref{fig:geo1} the mean particle current 
and the effective diffusion coefficient are depicted for different maximum channel widths.
Strikingly, the particle current $\langle \dot{x} \rangle$ decreases 
with increasing effective temperature $D$ \cite{Burada_CPC, Reguera_PRL}. 
This is in contrast to the transport over energetic potential barriers 
where thermal activation leads to an increasing particle current when 
the temperature is increased \cite{hanggithomas}. In our system, temperature dictates 
the height of the entropic barriers. With increasing temperature, 
the height of the barriers increases and the particle current decreases, cf. Fig.~{\ref{fig:geo1}}(a). 
In the deterministic limit, i.e., $D \rightarrow 0$ the average particle current 
equals to the applied bias $F$ irrespective of the maximum channel width. 
In this limit, the particles do not explore the side bags of the channel structure 
and move within a region defined by the width at the bottleneck. Consequently, 
the particles assume the velocity of a biased free particle. However, this observation is not captured 
by the analytics, as the influence of the winding of the structure is 
overestimated in the spatially dependent diffusion coefficient \cite{Burada_BioSy}.
 
As the maximum width increases, the area of the size bags increases. 
More available space in the orthogonal channel, however, leads to reduced 
particle current along the channel direction. Therefore, the particle current 
decreases with increasing maximum width, cf. Fig.~\ref{fig:geo1}(a). 
A similar reasoning applies for the increased enhancement of diffusion, 
which becomes visible in Fig.~\ref{fig:geo1}(b). 
The effective diffusion coefficient does not depend monotonously on 
the temperature $D$, and exhibits the effect of enhancement of diffusion \cite{Reguera_PRL}.

Overall, for small values of $w_\mathrm{max}$, i.e.,
smooth geometries the analytical description leads to better
results whereas it fails for large $w_\mathrm{max}$ and 
small $D$ range \cite{Burada_CPC, Burada_BioSy, Burada_PRE}.
Also the applicability of the analytical description depends
on the parameter we look at, i.e., whether it is the average
particle current (1st order) or the effective diffusion coefficient 
(2nd order).

\subsection{Constant-width-scaling}

\begin{figure}[t]
 \centering
 \includegraphics{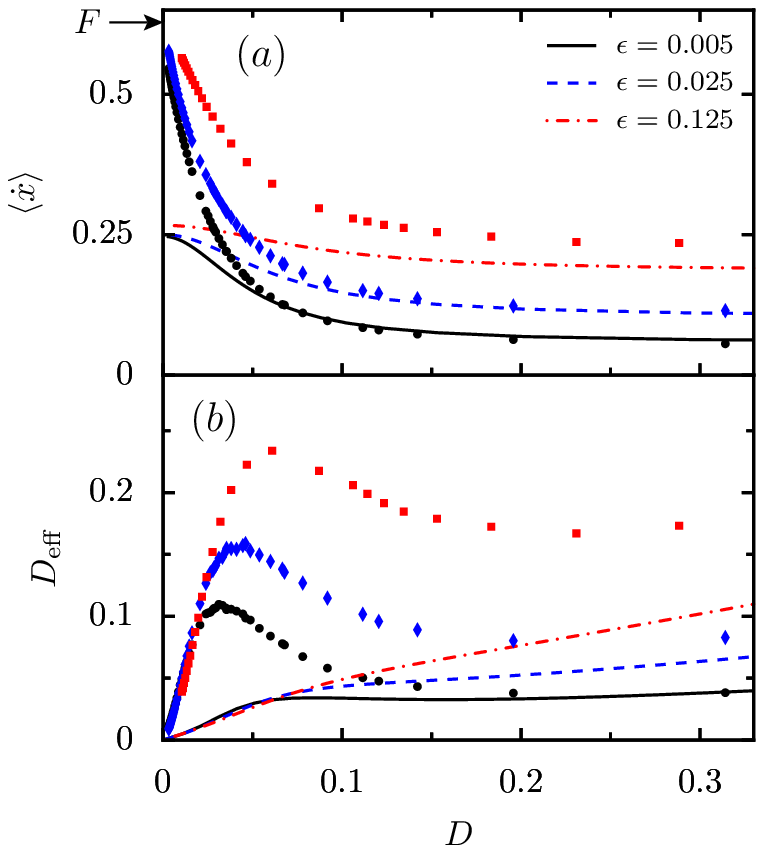}
 \caption{(Color online) Noise (temperature) dependence of the
   average particle current (a) and the effective diffusion 
   coefficient (b) for a symmetric two-dimensional channel 
   with the shape defined by Eq.~(\ref{eq:geometryus}) at the force 
   value $F = 0.628$ and for various values of the aspect ratio ($\epsilon$)
   of the geometry and for a constant maximum half-width
   $w_{\mathrm{max}} = 1$. Like in Fig.~\ref{fig:geo1} lines
   correspond to analytic results and symbols to numerical
   results. Moreover, the arrow shows the particle current for the
   deterministic limit.}
 \label{fig:geo2}
 \end{figure}

If we keep the maximum width constant and increase the $\epsilon$-value, 
the strength of the entropic nature of the system, i.e. the height of 
the entropic barrier in our system, is reduced.  
In Fig.~\ref{fig:geo2} we depict the dependence of the average particle 
current and the effective diffusion coefficient on the 
noise strength. Once more, the particle current exhibits this peculiar 
dependence on the noise level which is characteristic for the transport 
over entropic barriers. Namely, the particle current decreases with 
increasing noise level. In the deterministic limit, the particle 
current tends to the applied bias $F$  for all values of $\epsilon$, cf. Fig.~\ref{fig:geo2}(a).
Upon decreasing the aspect ratio $\epsilon$, the width of the 
bottleneck decreases and the effective height of the barrier increases. 
Consequently, the particle current decreases with decreasing $\epsilon$. 

Moreover, by decreasing the barrier height, i.e. increasing the $\epsilon$-value, 
the effect of the enhancement of diffusion becomes less significant, 
and in the limit of a flat channel, i.e. for $\epsilon \to 0$, there is linear, 
monotonic behavior of the effective diffusion coefficient, cf. Fig.~\ref{fig:geo2}(b).

Overall, for the constant-width-scaling, the  analytical description leads 
to better results for the geometries with very small aspect ratio. 
With aspect ratio $\epsilon$, the bending of the boundary function and, 
therefore, its derivative increases and the criteria for the applicability 
of the Fick-Jacobs approximation, i.e. $\omega'(x)\ll 1$, increasingly fails, cf. Fig.~\ref{fig:geo2}.

\subsection{Comparison of the different scalings}

The transport process depends on many factors such as the slope 
of the structure, the width of the channel at the bottleneck, 
the strength of the applied bias, and 
the thermal noise present in the system. The quality of the transport can be 
measured by the $Q$-factor which is introduced in Sec.~\ref{sec:transchar}. 
A large $Q$-factor means more randomness in the transport process.

\begin{figure}[t]
 \centering
 \includegraphics{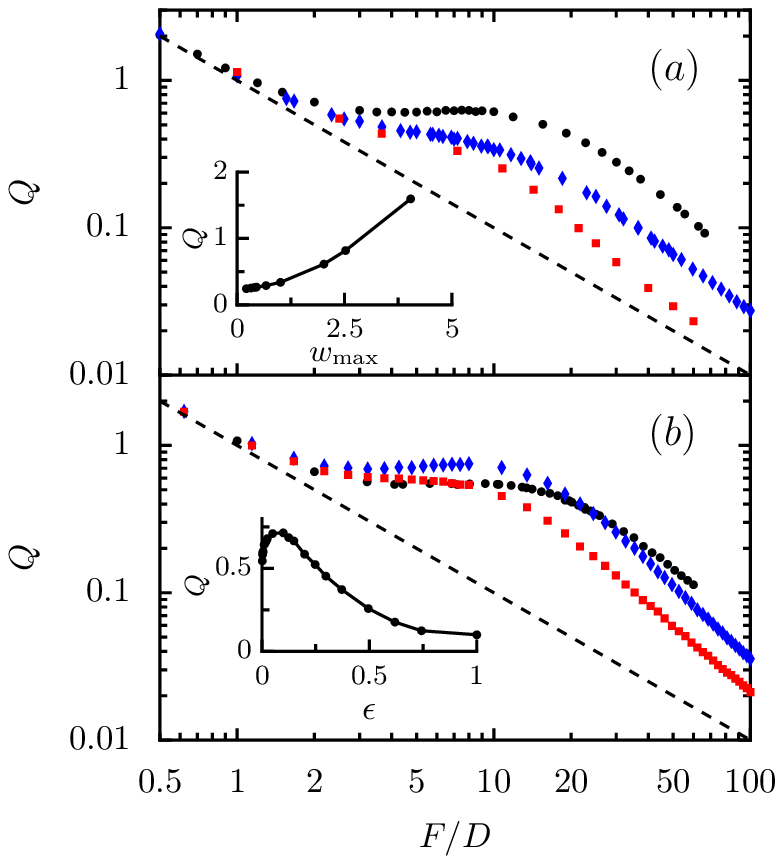}
 \centering
 \caption{(Color online) 
   The behavior of $Q$ factor as a function of the scaling parameter
   $F/D$ obtained from numerical simulations
   $(a)$ for the case of constant-ratio-scaling, where the aspect ratio 
   $\epsilon = 0.01$, and for various values of
   the maximal half-width $w_\mathrm{max}$. Circles: $w_\mathrm{max} = 2.02$;
   triangles: $w_\mathrm{max} = 2.02/2$; squares: $w_\mathrm{max} = 2.02/2\pi$.
   The inset in $(a)$ depicts the behavior of $Q$ as a function of $w_\mathrm{max}$ 
   at $F/D = 10$.
   $(b)$ for the case of constant-width-scaling, 
   where the maximal half-width $w_\mathrm{max} = 1$, and for various values of the 
   aspect ratio $\epsilon$. Circles: $\epsilon = 0.001$; triangles: $\epsilon = 0.075$;
   squares: $\epsilon = 0.03$. 
   The inset in $(b)$ depicts the behavior of $Q$ as a function of $\epsilon$ at $F/D = 10$.
   The dashed line in both the figures correspond to $Q_{\mathrm{free}} = 1/(F/D)$.}
  \label{fig:Q-diffusion}
\end{figure}

Fig.~\ref{fig:Q-diffusion} shows the dependence of $Q$
for the two different scalings. Interestingly, with increasing maximum 
width of the channel, while keeping the aspect ratio of channel constant, 
the $Q$-factor increases. Means, the transport becomes more noisy for larger
channel widths, cf. the behavior of $Q$-factor 
in the inset of Fig.~\ref{fig:Q-diffusion}(a). 
The maximum of the $Q$-factor for wider channels is accompanied
by a increased enhancement of diffusion on one side and on the other side 
the reduced particle current, cf. Sec.~\ref{sec:crscaling} and, in particular, Fig.~\ref{fig:geo1}. 

Within the constant-width-scaling, smaller $\epsilon$ leads 
to a decrease of the height of the entropic barrier.
Hence, the transport should be better in the sense of a smaller $Q$-factor.
However, we found that for rather small and fixed values 
of $w_\mathrm{max}$ an increase of $\epsilon$ could 
spoil the quality of transport by  increasing $Q$, cf. the behavior of 
the $Q$-factor in the inset of Fig.~\ref{fig:Q-diffusion}(b). 
Thus, by varying the aspect ratio $\epsilon$ of the geometry one can 
encounter different transport regimes.
If the particular scaling parameter $F/D$ is given, one can tune the quality of transport
by choosing an optimal geometry, i.e., an optimal $\epsilon$-value.
On the other hand, if the structure is given, the scaling parameter 
$F/D$ may be used to control the transport. 
An optimal transport in the sense of a smallest $Q$-factor could always be
achieved at either very small or very large scaling parameters $F/D$, cf. Fig.~\ref{fig:Q-diffusion}.

\section {Conclusions}

In this work we analyzed the transport of Brownian particles through 
channels with periodically varying width, which exhibit narrow bottlenecks.  
The effect of confinement can be recast in terms of an entropic potential, and
the dynamics of the system can be effectively described by means of
the Fick-Jacobs equation with a spatially dependent diffusion coefficient.
We observe peculiar transport phenomena, which are characteristic 
for the transport over the entropic potential barriers. 
The transport characteristics like the average particle current and 
the effective diffusion coefficient sensitively depend on the geometry of the channel. 
Depending on the geometric parameters we have considered two scaling regimes: 
the {\itshape constant-width-scaling}, for which we keep the maximum width constant, 
and the {\itshape constant-ratio-scaling} where the ratio of the maximum and 
the minimum width of the channel is kept fixed. 
For the former case, the diffusion process varies upon changing the 
aspect ratio of the geometry.
There is a critical bottleneck opening related to a maximum
effective diffusion coefficient.
This feature is more visible in highly confined geometries, i.e., for smaller 
$w_{\mathrm{max}}$ values.
In both geometric scalings, during the crossover region from a purely
energetic to entropy dominated regime, 
optimal transport could be observed,
suggesting that by increasing the noise strength $D$
the system may exhibit better transport, i.e., smaller $Q-$ factor.
Thus, tuning the parameters the aspect ratio $\epsilon$, 
the maximal half-width $w_{\mathrm{max}}$, and $D$
one can effectively regulate the transport characteristics in these 
confined geometries.
However, in general, controlling the width ratio $\epsilon$ and 
the maximal half-width $w_{\mathrm{max}}$
of the geometry may not be feasible, but, tuning the scaling
parameter $F/D$ one can arrive at an optimal transport regime.

\subsection*{ACKNOWLEDGMENTS}

This work was made possible thanks to the financial support by: 
the Max-Planck society,
the Volkswagen Foundation (project I/80424); 
the Deut\-sche Forschungsgemeinschaft (DFG) via project no. 1517/26-2 
and via the collaborative research center, SFB-486, project A10; 
Nanosystems Initiative Munich (NIM).


\begin{thebibliography}{00}
\bibitem{Burada_CPC}
  P. S. Burada, P. H\"anggi, F. Marchesoni, G. Schmid, P. Talkner,
  {\it ChemPhysChem}, \textbf{10}, 45 (2009).

\bibitem{Reimann_PRL}
P. Reimann, C. Van den Broeck, H. Linke, P. H\"anggi, J. M. Rub\'i,
A. P\'erez-Madrid, \emph{Phys. Rev. Lett.}, {\bfseries 87},
010602 (2001); \emph{Phys. Rev. E} {\bfseries 65}, 031104 (2002).

\bibitem{Lindner_FNL}
B. Lindner,  M. Kostur, L. Schimansky-Geier, 
{\it Fluctuation and  Noise Letters} {\bfseries 1},  R25 (2001).

\bibitem{Reguera_PRL}D. Reguera, G. Schmid, P.S. Burada,
  J.M. Rub\'i, P. Reimann, P. H\"anggi,
  Phys. Rev. Lett. \textbf{96}, 130603 (2006).

\bibitem{hille}B. Hille, {\it Ion Channels of Excitable Membranes}
  (Sinauer, Sunderland, 2001).

\bibitem{liu}
  L. Liu, P. Li, S.A. Asher,
  {\it Nature} \textbf{397}, 141 (1999).

\bibitem{siwy}
  Z. Siwy, I.D. Kosinska, A. Fulinski, C.R. Martin,
  {\it Phys. Rev. Lett.} \textbf{94}, 048102 (2005).

\bibitem{berzhkovski}
  A.M. Berezhkovskii, S.M. Bezrukov,
  {\it Biophys. J.} \textbf{88}, L17(2005).

\bibitem{zeolites}
  R.M. Barrer,
  {\it Zeolites an Clay Minerals as Sorbents and Molecular Sieves}
  (Academic Press, London, 1978).

\bibitem{Chou}
  T. Chou, D. Lohse,
  {\it Phys. Rev. Lett.} \textbf{82}, 3552 (1999).

\bibitem{kettner}
  C. Kettner, P. Reimann, P. H\"anggi, F. M\"uller,
  {\it Phys. Rev. E}  \textbf{61}, 312 (2000).

\bibitem{muller}
  S. Matthias, F. M\"uller,
  {\it Nature} \textbf{424}, 53 (2003).

\bibitem{Austin}
  W.D. Volkmuth, R.H. Austin,
  {\it DNA electrophoresis in microlithographic arrays},
  Nature \textbf{358}, 600 (1992).

\bibitem{Nixon}
  G.I. Nixon, G.W. Slater,
  {\it J. Chem. Phys.} {\bfseries 117}, 4042 (2002).

\bibitem{Chang}
  R. Chang, A. Yethiraj,
  {\it Phys. Rev. Lett.} {\bfseries 96}, 107802 (2006).

\bibitem{Kosinska}
  I.D. Kosinska, I. Goychuk, M. Kostur, G. Schmid,  P. H\"anggi,
  {\it Phys. Rev. E} \textbf{77}, 031131 (2008).

\bibitem{revmodphys}
P. H\"anggi, F. Marchesoni, \emph{Rev. Mod. Phys.} {\bfseries 81},
387 (2009). 

\bibitem{BM}
P. H\"anggi, F. Marchesoni,  F. Nori, \emph{Ann. Physik (Berlin)}
\textbf{14}, 51 (2005).


\bibitem{PT}
R. D. Astumian, P. H\"anggi, \emph{Physics Today} \textbf{55} (11), 33 (2002).

\bibitem{RH}
P. Reimann, P. H\"anggi, \emph{Appl. Physics A} \textbf{75}, 169 (2002).

\bibitem{Keyser-RSIns}
  U.F. Keyser, J. van der Does, C. Dekker, N.H. Dekker,
  {\it Rev. Sci. Instrum.} \textbf{77}, 105105 (2006).

\bibitem{Han}
  J. Han, H.G. Craighead,
  {\it Science} \textbf{288}, 1026 (2000).

\bibitem{Keyser} 
  U.F. Keyser, B.N. Koeleman, S. Van Dorp, D. Krapf,
  R.M.M. Smeets, S.G. Lemay, N.H. Dekker, C. Dekker, 
 {\it Nature Physics} {\bfseries 2}, 473 (2006).


\bibitem{Aksimentiev}
  A. Aksimentiev, B.J. Heng, G. Timp, K. Schulten,
  {\it Biophys. J.} \textbf{87}, 2086 (2004).


\bibitem{Heng_1}
  J.B. Heng, C. Ho, T. Kim, R. Timp, A. Aksimentiev, Y.V. Grinkova,
  S. Sligar, K. Schulten, G. Timp,
  {\it Biophys. J.} \textbf{87}, 2905 (2004).


\bibitem{Heng_BPJ}
  J. B. Heng, A. Aksimentiev, C. Ho, P. Marks, Y.V. Grinkova,
  S. Sligar, K. Schulten, G. Timp,
  {\it Biophys. J.} \textbf{90}, 1098 (2006).

\bibitem{Gerland}
  U. Gerland, R. Bundschuh, T. Hwa, 
  {\it Phys. Biol.} {\bfseries 1}, 19 (2004).


\bibitem{Bundschuh}
  R. Bundschuh, U. Gerland, 
  {\it Phys. Rev. Lett.} {\bfseries 95}, 208104 (2005).

\bibitem{Kosinska_APP}
  I. D. Kosinska, I. Goychuk, M. Kostur, G. Schmid, P. H\"anggi,
  {\it Acta Physica Polonica B} \textbf{39}, 1137 (2008).


 \bibitem{Brun}
   L. Brun, M. Pastoriza-Gallego, G. Oukhaled, J. Math\'e,
   L. Bacri, L. Auvray, J. Pelta,
   {\it Phys. Rev. Lett.} \textbf{100}, 158302 (2008).


 \bibitem{Kasianowicz}
   J.J. Kasianowicz, E. Brandin, D. Branton, D.W. Deamer,
   {\it Proc. Natl. Acad. Sci.} \textbf{93}, 13770 (1996).


\bibitem{Choi}
  Y. Choi, A. Mecke, B.G. Orr, M.M. Banaszak Holl, J.R. Baker (Jr.),
  {\it Nano Lett.} \textbf{4} 497, (2004).


\bibitem{Purcell} E.M. Purcell, 
  \emph{Am. J. Phys.} {\bf 45}, 3 (1977).


\bibitem{hanggithomas} P. H\"anggi, H. Thomas, 
  {\it Phys. Rep.} \textbf{88}, 207 (1982).


\bibitem{Risken}
  H. Risken, {\it The Fokker-Planck equation}, 2nd ed. (Springer, Berlin, 1989).


\bibitem{Jacobs}M.H. Jacobs, Diffusion Processes (Springer, New York, 1967).


\bibitem{Zwanzig}R. Zwanzig, \emph{J. Phys. Chem}, \textbf{96}, 3926 (1992).


\bibitem{Reguera_PRE}D. Reguera, J.M. Rubi,
\emph{Phys. Rev. E} \textbf{64}, 061106 (2001).


\bibitem{Burada_PRE} 
  P.S. Burada, G. Schmid, D. Reguera, J.M. Rub\'i, P. H\"anggi, 
  \emph{Phys. Rev. E} \textbf{75}, 051111 (2007).

  
\bibitem{Burada_BioSy} 
  P.S. Burada, G. Schmid, P. Talkner, P. H\"anggi, D. Reguera, J.M. Rub\'i, 
  \emph{BioSystems} \textbf{93}, 16 (2008).

\bibitem{Kalinay_PRE} P. Kalinay, J.K. Percus, Phys. Rev. E
  \textbf{74}, 041203 (2006). 
	
\bibitem{Kalinay_PRE09}
 P. Kalinay, \emph{Phys. Rev. E} {\bfseries 80}, 031106 (2009).
  
\bibitem{Burada_PTRSA}
P.S. Burada, G. Schmid, P. H\"anggi, {\itshape Phil. Trans. R. Soc. A} {\bfseries 367}, 3157 (2009).
\end{thebibliography}
\end{document}